\documentclass[fleqn,usenatbib]{mnras}

\usepackage{newtxtext,newtxmath}
\usepackage[T1]{fontenc}
\DeclareRobustCommand{\VAN}[3]{#2}
\let\VANthebibliography\thebibliography
\def\thebibliography{\DeclareRobustCommand{\VAN}[3]{##3}\VANthebibliography}
\usepackage{graphicx}	
\usepackage{amsmath}	

\title[UDGs and the SMHR]{Do Ultra Diffuse Galaxies with Rich Globular Clusters Systems have Overly Massive Halos?}

\author[D. A. Forbes and J. Gannon]{
Duncan A. Forbes, $^{1}$\thanks{E-mail: dforbes@swin.edu.au}
Jonah Gannon, $^{1}$ 
\\
$^{1}$ Centre for Astrophysics \& Supercomputing, Swinburne University, Hawthorn, VIC 3122, Australia
}

\date{Accepted XXX. Received YYY; in original form ZZZ}

\pubyear{2023}

\begin{document}
\label{firstpage}
\pagerange{\pageref{firstpage}--\pageref{lastpage}}
\maketitle

\begin{abstract}

Some Ultra Diffuse Galaxies (UDGs) appear to host exceptionally rich globular cluster (GC) systems compared to normal galaxies of the same stellar mass. After re-examing these claims, we focus on a small sample of UDGs from the literature that have {\it both} rich GC systems (N$_{GC}$ $> 20$) and a measured galaxy velocity dispersion. We find that UDGs with more GCs have higher dynamical masses and that GC-rich UDGs are dark matter dominated within their half-light radii. We extrapolate these dynamical masses to derive total halo masses assuming cuspy and cored mass profiles. 
We find reasonable agreement between halo masses derived from GC numbers (assuming the GC number - halo mass relation) and from cored halo profiles. This suggests that GC-rich UDGs do {\it not} follow the standard stellar mass -- halo mass relation, occupying overly massive cored halos for their stellar mass. 
A similar process   
to that invoked for some Local Group dwarfs, of early quenching,  may result in GC-rich UDGs that have failed to form the expected mass of stars in a given halo (and thus giving the appearance of overly an massive halo). Simulations that correctly reproduce the known properties of GC systems associated with UDGs are needed.

\end{abstract}


\section{Introduction} \label{sec:intro}

In the Lambda Cold Dark Matter ($\Lambda$CDM) paradigm the evolution of a galaxy is intimately tied to its surrounding halo of dark matter. The halos are thought to control the availability of gas which dissipates energy and converts into stars as the gas falls into the centre of the halo, thus building a galaxy. Subsequent mergers of halos (and gas plus stars) contribute to mass growth over cosmic time. 
Within a $\Lambda$CDM cosmology, numerical simulations and theoretical models seek to describe the detailed properties of 
galaxies and their host halos. While these have had many successes at large scales and at high masses, they are also in tension with some observations, particularly in the low galaxy mass regime (see review by 
\citealt{2022NatAs...6..897S}). 
Such tensions have led to alternatives to the standard LCDM paradigm being proposed (e.g. 
\citealt{2012IJMPD..2130003K}).

The galaxy-halo connection can be probed by the stellar mass--halo mass relation (SMHR). This relation describes the mass of a galaxy in stars to the total mass of its host halo. Its shape thus reveals the efficiency of star formation in a given halo, with the maximum efficiency occurring at a halo mass of log M$_{h}$ $\sim$ 12 M$_{\odot}$.  
As summarised by 
\cite{2020MNRAS.499.4748M}, 
low mass galaxies are generally expected to assemble earlier than high mass galaxies, have higher fractions of in-situ formed stars and take longer to form most of their stellar mass.  Whereas high mass galaxies form their stars rapidly and continue to assemble mass due to the later accretion of ex-situ material (e.g.  \citealt{2008MNRAS.389..567C}).
Low mass passive galaxies tend to have lower halo masses than actively star forming galaxies. 
Although, it is important to note that at higher stellar masses, this trend inverts. 
Galaxies may also be subject to tidal stripping, 
resulting in a galaxy with a reduced halo mass, at a given stellar mass,  since DM material is preferentially removed before stars 
(\citealt{2023MNRAS.523.6020W}). 
Thus passive galaxies of low stellar mass and/or those tidally stripped might be expected to have {\it lower} mean halo masses at the stellar masses typical of UDGs. 
We note that a warm DM universe would also tend to scatter galaxies to lower halo masses for a given stellar mass 
(\citealt{2016MNRAS.459.2573R}).
For a review of the SMHR,  
and recent developments, see 
\citet{2018ARA&A..56..435W} and 
\cite{2023ApJ...956....6D}.

The importance of ultra diffuse galaxies (UDGs) in the context of the SHMR, was first highlighted by 
\citet{2015ApJ...798L..45V} 
who used the novel Dragonfly Telephoto array to identify a population of them in the Coma cluster. 
They proposed 
that they may lie within halos more massive than predicted from the standard SMHR in order to protect them from the harsh cluster environment. 
Subsequent support for this idea comes from the dynamical studies of UDGs
which indicate that they are DM-dominated in their inner regions compared to normal galaxies of the same stellar mass (e.g. 
\citealt{2018ApJ...856L..31T}; \citealt{2022MNRAS.510..946G}). 
Although of low stellar 
mass (log M$_{\ast}$ $\sim$ 8), UDGs are still too massive to have their star formation regulated by reionisation. They are however in the mass regime to be strongly affected by supernova feedback
(e.g. \citealt{2017MNRAS.466L...1D}).

UDGs are defined, somewhat arbitrarily, 
as having effective (half projected light) sizes R$_e$ $\ge$ 1.5 kpc and low surface brightnesses of $\mu_{g,0}$ $\ge$ 24 mag per sq. arcsec. 
This definition implies UDGs  have low stellar mass densities, and given a small range in sizes from 1.5 kpc to the largest UDGs known, a narrow range in stellar mass. It is unknown whether such galaxies reside in normal or overly-massive halos for their total stellar mass.
UDGs may represent around 5\% of galaxies in groups and clusters 
(e.g. 
\citealt{2017A&A...608A.142V}; \citealt{2022arXiv221014994L}). 
Thus UDGs are not especially rare and have been found in all environments, from the field to dense clusters, with  
their absolute number scaling  with the mass of the environment (e.g. 
\citealt{2017A&A...607A..79V}).
UDGs located in clusters tend to be red and passive, and in some cases, rich in globular clusters (e.g. see 
\citealt{2020MNRAS.492.4874F} 
for a compilation of GC counts around UDGs located in the Coma cluster).  
Some literature studies focusing on UDGs have been extended beyond the original definition to slightly smaller and/or brighter galaxies (e.g.
\citealt{2022arXiv221014994L}). 
Here we name these  nearly-UDGs NUDGES, since they nudge up against the standard definition of a UDG.

Only one UDG to date, DF44, a passive galaxy in the Coma cluster, has an extended radial kinematic profile and associated mass modelling to infer a total halo mass 
(\citealt{2019ApJ...880...91V}).
A cored halo profile with isotropic orbits provided a good fit to the kinematic data and was preferred over a cuspy profile that required strongly tangential orbits by 
\cite{2019ApJ...880...91V}.
With a cored halo the resulting total halo mass indicates 
that DF44 occupies an overly massive halo for its stellar mass. 
This observation required 17 hours of on-source Keck telescope time and so is unlikely to be available for a large sample of UDGs in the near future (we note that radial kinematics also exists for the dark-matter free UDG NGC1052\_DF2; \cite{2019A&A...625A..76E}).
While enclosed dynamical masses from stellar kinematics within the effective radius have been measured for a dozen UDGs (see 
\citealt{2023MNRAS.518.3653G} 
for an overview), deriving total halo masses from these data requires an assumption of the (unknown) mass profile.

Another approach for estimating total halo masses is to use the empirical scaling with the total number of GCs, or their system mass (see
\citealt{2009MNRAS.392L...1S}; \citealt{2017ApJ...836...67H};  \citealt{Burkert2020}).
This near-linear relation holds  for all galaxy types with small scatter over a wide range of galaxy masses. 
\cite{2017MNRAS.472.3120B}
described this relation as ``... a [natural] consequence of the mass assembly of CDM halos.."
However, while this relation covers the mass regime of UDGs it is unknown whether UDGs actually obey the relation or not. 
If UDGs do obey the relation, then those UDGs with particularly rich GC systems occupy overly massive halos and 
they do not obey the expectations of the standard SMHR seen for normal galaxies. In the case of Coma cluster UDGs, more than half of those observed may be GC-rich (\citealt{2020MNRAS.492.4874F}) and thus may have overly massive halos. 
Alternatively, if GC-rich UDGs obey the SMHR then they can not follow the GC number -- halo mass relation for normal galaxies. {\it Either outcome is interesting and reinforces the extreme nature of UDGs and their GC systems}. 

Another motivation to study UDGs with rich GC systems comes from the work of \cite{2022MNRAS.517.2231B}.
This study found GC-rich and GC-poor UDGs to reveal different locations in stellar mass-metallicity space. In particular, GC-rich UDGs have on average much lower stellar metallicities, at a given galaxy stellar mass, than GC poor UDGs or classical dwarf galaxies. 

Here we investigate the important question of whether GC-rich UDGs have overly massive halos for their stellar mass.  
First we briefly revisit the GC number -- halo mass relation (which we refer to as the GNHR). 
We then ask if the claims of rich GC systems can be believed given some conflicting results in the literature.  
For a sample of UDGs with both a rich GC system and a velocity dispersion measurement, we investigate whether the number of GCs scales with dynamical mass.  
We compare halo masses derived from different methods for our sample of GC-rich UDGs. We also compare to the standard stellar mass -- halo mass relation (SMHR). Finally, we discuss our results and the future work that is needed to make progress. 

\begin{figure}
	\includegraphics[width=0.8\linewidth, angle=-90]{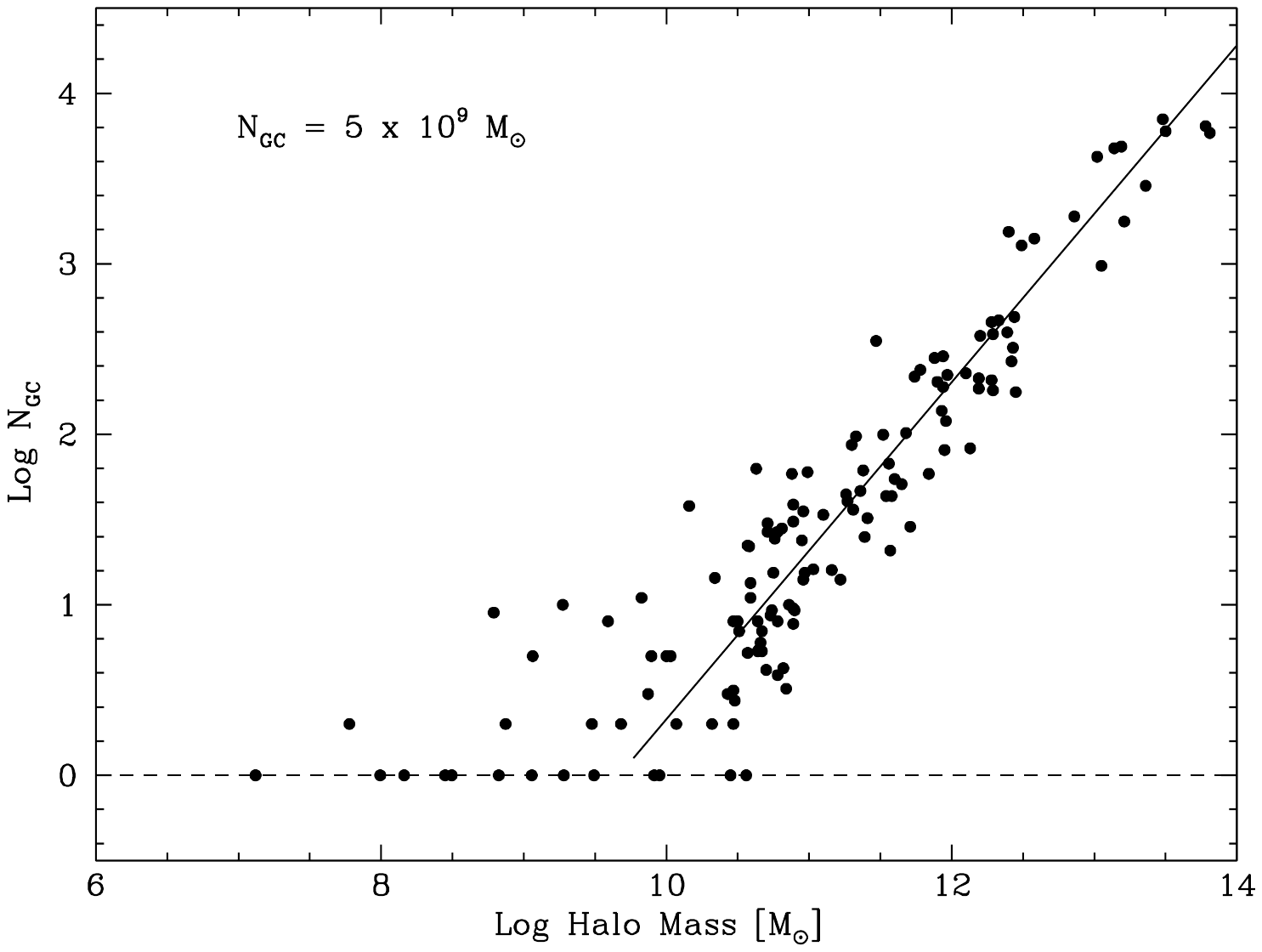}
	\caption{Number of globular clusters vs galaxy halo mass scaling relation (GNHR). Symbols show normal galaxies (i.e. non-UDGs)
from \protect\cite{2009MNRAS.392L...1S} 
Symbols with a value of zero (dashed line) indicate galaxies with no GCs. The solid line shows the scaling relation fit by 
\protect\cite{Burkert2020},
  which is consistent with a linear relation of 1 GC equivalent to a halo mass of 5 $\times$ 10$^9$ M$_{\odot}$. Below $\sim$15 GCs the predictive power of the relation is much reduced.}
\end{figure}

\section{GC Number -- Halo Mass Relation}

The scaling relation between GC number (or GC system mass) and total halo mass (GNHR) is a near linear relation over many orders of magnitude 
(\citealt{Burkert2020};
hereafter BF20). It is consistent with a simple scaling for each GC to be associated with a halo of mass 5 $\times$ 10$^9$ M$_{\odot}$ as shown by the solid line in Fig. 1. The scatter in the relation generally dominates over the measurement uncertainty in the GC counts of a given system. 
In this work halo mass is taken to be the mass within the radius corresponding to a mean density 200 times the critical density of the Universe at z = 0. When masses are quoted as log values, solar masses should be assumed.
BF20 measured  the scatter to be around 0.25 dex at a halo mass of log M$_{h}$ = 12 (corresponding to 200 GCs) increasing to 0.3 dex at log M$_{h}$ = 11 (corresponding to 20 GCs). Below $\sim$15 GCs the relation has considerably higher scatter and loses its halo mass predictive power for individual UDGs. However, it appears to continue to also hold to lower GC numbers in a statistical sense 
\citep{2023ApJS..267...27Z}.
The relation holds for 
all types of normal galaxies, from giants to dwarfs, from early to late types 
\citep{2009MNRAS.392L...1S}. 
The latest simulations that include GCs (e.g. 
\citealt{2021MNRAS.505.5815V}; \citealt{2023MNRAS.522.5638C};
\citealt{2023MNRAS.518.2453D};
\citealt{2023arXiv230702530D})
can reproduce the near linear slope. 
It is currently  unknown whether, or not, UDGs follow the GNHR for normal galaxies. 

\section{Are the rich GC systems of some UDGs to be believed?}

The reported GC systems for UDGs reveal a large range of richness, from those hosting no GCs to those that host rich GC systems (many 10s) for their stellar mass. 
For example, 
\cite{2020MNRAS.492.4874F}
combined several literature HST imaging studies of $\sim$50 Coma cluster UDGs to show that over half host rich GC systems. In particular, they revealed
GC specific frequencies (S$_N$ = N$_{GC}$ $\times$ 10$^{0.4 (M_V +15)}$) 
up to $\sim$100 and ratios of GC system mass to that of the host galaxy stellar mass of up to $\sim$10\%. On average, Coma cluster UDGs appear to host up to several times more GCs than classical dwarf galaxies of comparable stellar mass. 

Recently, the number of GCs in five Coma cluster UDGs has been questioned by 
\cite{2022MNRAS.511.4633S} (hereafter S22).
Could these five, and potentially other previous literature estimates of GC systems associated with UDGs, be in error? Below we discuss the recent controversy surrounding the GC counts associated with five Coma cluster UDGs.  
But before discussing these specific cases, it is worthwhile to summarise how the total number of GCs around a given host galaxy is derived from imaging. 

While there is no universally agreed upon approach to estimating the number of GCs associated with a galaxy, most imaging studies take a similar approach. Initially, an automatic detection algorithm is run. This is refined using various known properties of GCs as might be expected for a distant galaxy (e.g. 
angular size, apparent magnitude, colour if several filters are available, etc). For ground-based imaging, GCs with half-light sizes of 2--10pc, are generally unresolved beyond the Local Group. However, imaging with HST may partially resolve GCs out to distances of $\sim$30 Mpc, thus providing more confidence in their validity. After selection criteria are applied, contamination (from foreground stars,  background galaxies and perhaps intracluster GCs) can be estimated statistically by examining selected object detections that are well separated from the galaxy on the sky. 

GC samples, determined in this manner, are likely to be incomplete in terms of both radial coverage and magnitude. The former can be corrected by integrating the radial profile to a large radius, once a background level of constant density (i.e. contaminants) is removed. This then gives the total number of GCs down to some magnitude limit. 
The latter can be corrected for 
by understanding the detection rate of mock GCs, as a function of magnitude, that have been inserted into the image. 
Once the universal peak in the GC luminosity function is reached, the number of GCs brighter than the peak can be simply doubled to estimate the number of missing faint GCs. Thus the total number of GCs over all radii and all luminosities can be derived. Strictly speaking, imaging, in the absence of radial velocities, only gives GC candidates but a radial profile of GC density that falls off from a galaxy centre is a strong indicator of association. 
A recent example of this process is described in 
\cite{2022MNRAS.509..180L}. 

\begin{table}
  \centering
  \begin{tabular}{l|c|c|c|c|c}
  \hline
      & vD17 & L18 & S22 & R$_{GC}$/R$_e$ & n\\
     \hline 
     \medskip
DF7 & -- & 39.1$\pm$23.8 & 22$^{+5}_{-7}$ & 0.98$^{+0.50}_{-0.38}$ & 1.12$^{+1.86}_{-0.88}$\\
\medskip
DF8 & -- & 42.6$\pm$23.9 & 10$^{+5}_{-8}$ & 0.68$^{+0.52}_{-0.38}$ & 0.36$^{+3.50}_{-0.16}$\\
\medskip
DF17 & 25$\pm$11 & 28$\pm$14 & 26$^{+17}_{-7}$ & 1.54$^{+0.28}_{-0.30}$ & 0.20$^{+0.34}_{-0.00}$\\
\medskip
DF44 & 76$\pm18$ & -- & 20$^{+6}_{-5}$ & 0.78$^{+0.44}_{-0.30}$ & 0.94$^{+1.74}_{-0.54}$\\
\medskip
DFX1 & 63$\pm$17 & -- & 17$^{+5}_{-6}$ & 0.80$^{+0.34}_{-0.26}$ & 1.06$^{+1.20}_{-0.54}$\\
     \hline
  \end{tabular}
  \caption{Comparison of vD17, L18 and S22 GC counts for 5 Coma UDGs in common. The last 2 columns are from S22, i.e. size of GC system relative to galaxy effective radius and Sersic n parameter from a fit to the GC system. Note: 
  \protect\cite{2016ApJ...819L..20B} derived a total of 27 $\pm$5 GCs for DF17. }
  \label{tab:fitparams}
\end{table}

Returning to the question of whether some GC count estimates for UDGs are in error, we 
discuss the controversy around several Coma cluster UDGs. S22 studied the GC systems of 6 Coma UDGs using HST. Of these, five  are in common with the previously published works by 
\cite{2017ApJ...844L..11V} (hereafter vD17) 
and \cite{2018ApJ...862...82L} (hereafter L18), 
who also used HST imaging to identify GCs. 
 The GC numbers found by S22 would, at face value, suggest that those of both L18 and vD17 were  overestimated by a factor of up to 4--5. 

\begin{figure}
	\includegraphics[width=0.8\linewidth, angle=-90]{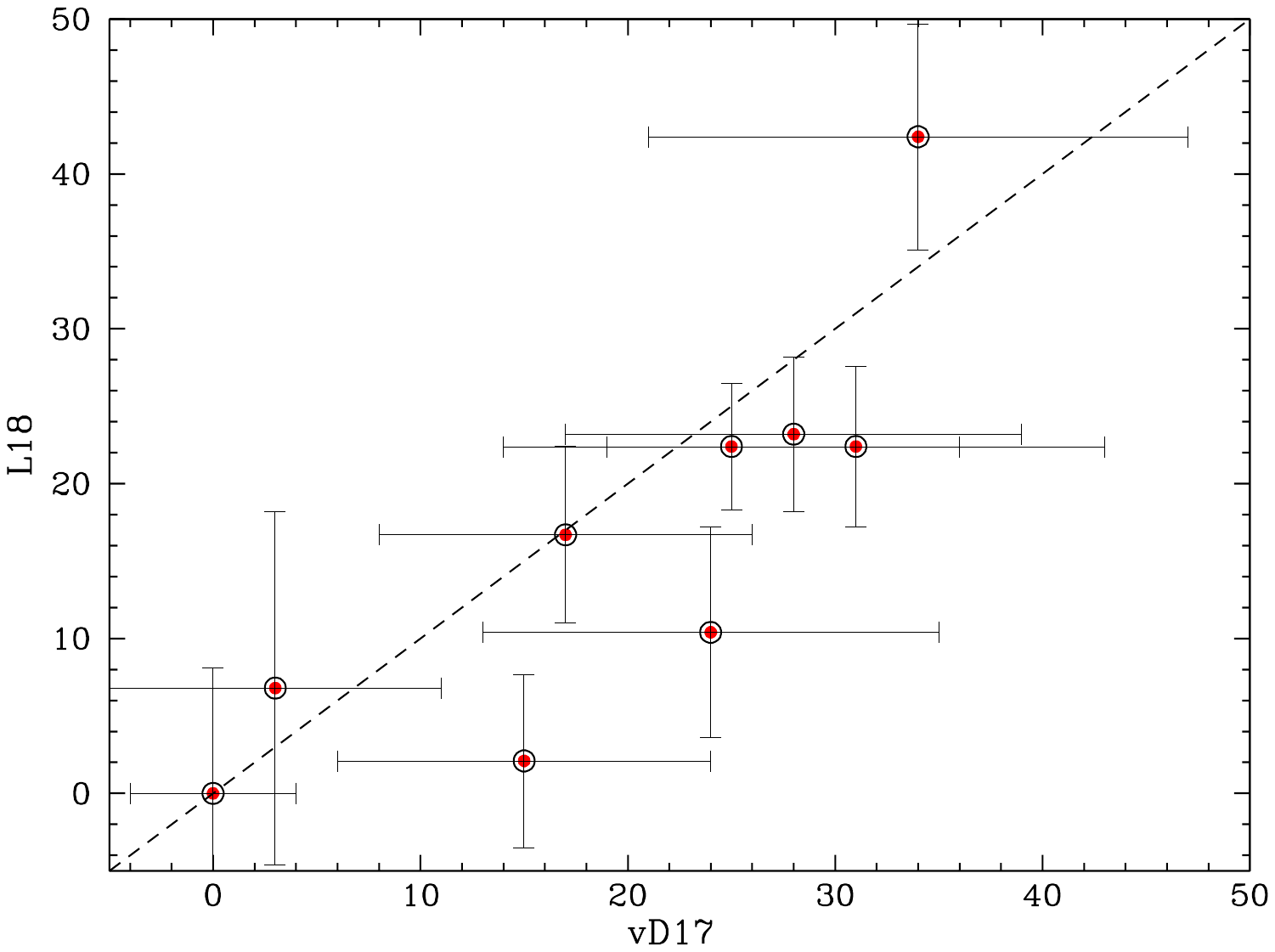}
	\caption{Comparison of literature 
  estimates for GC counts around Coma cluster UDGs. GC counts  from L18 are shown against from GC counts from vD17 for the 9 UDGs in common. These studies both used GC system to galaxy sizes of R$_{GC}$/R$_e$ = 1.5. Within their joint uncertainties their GC numbers are consistent with each other. 
  }

\end{figure}
  
\begin{figure}
	\includegraphics[width=0.8\linewidth, angle=-90]{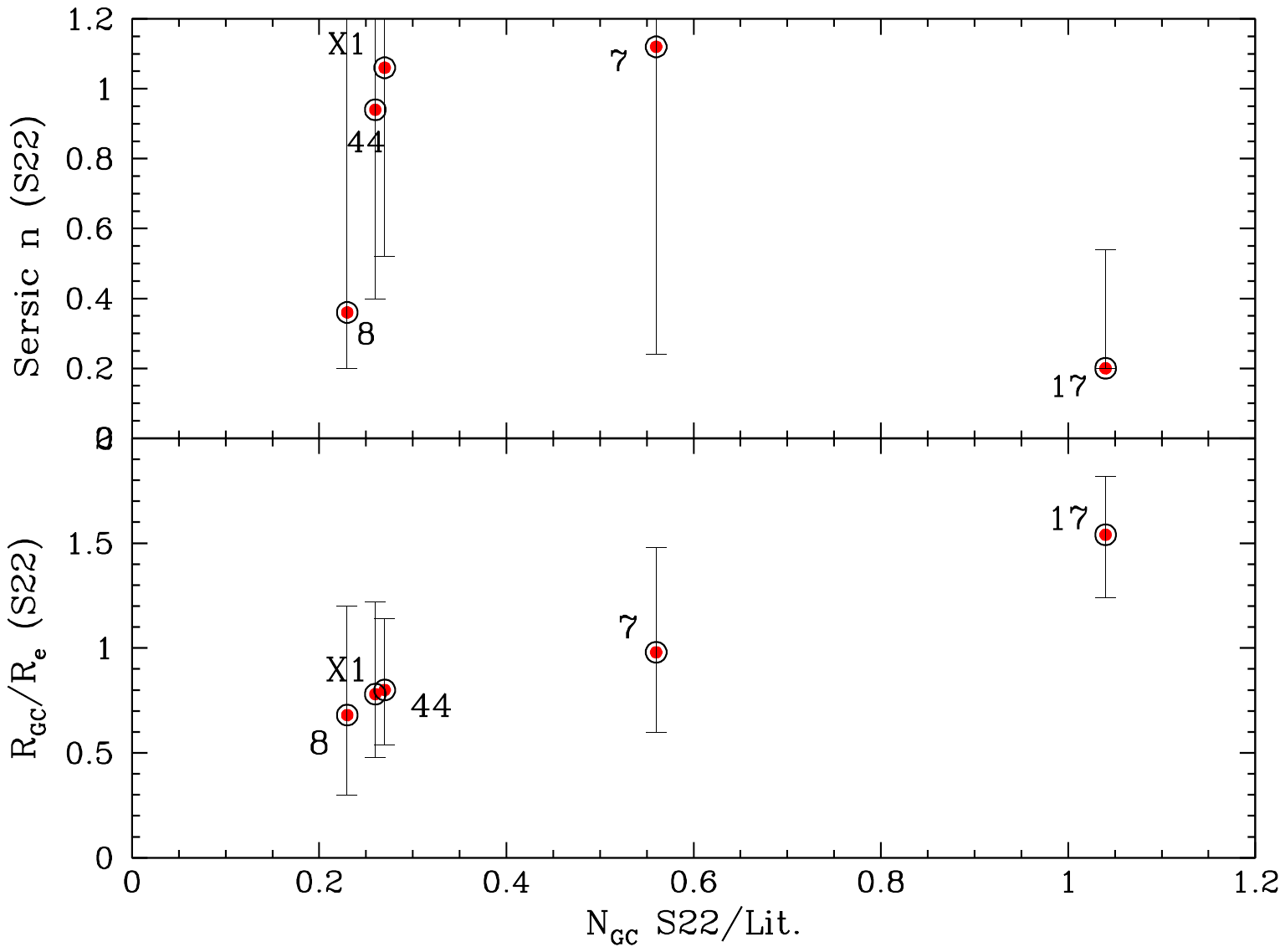}
	\caption{Comparison of literature 
  estimates (i.e. vD17 or L18) for GC counts around Coma cluster UDGs with the study of S22. The UDGs are labelled in each panel. For DF17 we use the vd17 value.  See Table 1 for error bars.   
 {\it Upper panel} shows the GC count ratio vs Sersic n fit to the GC system by S22. 
  {\it Lower panel} shows the GC count ratio 
 vs GC system size relative to galaxy effective radius fit by S22.  The x axis in both panels is number of GCs found by S22 relative to the literature studies of L18 and vD17. The lower panel shows that this ratio depends strongly on the GC system size determined by S22.  }
\end{figure}

In Table 1 we list the published data for the five Coma UDGs. We include GC counts from vD17, L18 and S22. We also include the GC system size (i.e. the half number radius relative to the galaxy effective radius) and the Sersic n parameter from a fit to the GC system radial density profile taken directly from table 3 of S22. 

We start by showing in Fig. 2 a comparison of the GC counts from L18 with those of vD17 for the 9 Coma cluster galaxies in common. We have set the L18 value for Y122 to 0 (i.e. no GC system) as L18 determined its contaminants to exceed its GC candidates. 
 Although VD17 finds half a dozen more GCs on average than L18, the GC counts between the two independent studies are generally consistent with each other within their joint uncertainties. This gives us some confidence in the GC counts from both studies.  
 
Fig. 3 compares the total number of GCs found by S22 divided by either the GC number quoted by vD17 or L18 versus the GC system size (lower panel) and the Sersic n parameter (upper panel) for the GC system radial profile.  The steepness of the Sersic profile as quantified by n (e.g. n = 1 is an exponential disk-like profile) does not seem to be the reason for the difference. However, the size of the GC system, as determined by S22, does appear to be a strong driver of the differences in the GC counts from S22 compared to those of the other two studies. 

The five galaxies include DF17 where S22 finds a similar GC count as vD17 and L18, where GC system size is around 1.5 times the galaxy effective radius in all three studies.
DF17 has also been studied by 
\cite{2016ApJ...819L..20B}
finding 27 $\pm$ 5. Thus the 4 studies are in
remarkable agreement, with a rounded average of 27 GCs with an error on the mean of less than 1 GC.  
In contrast, S22 finds only 23\% of the GCs found by L18 for DF8. Here the GC system size determined by S22 is the smallest at R$_{GC}/R_e$ = 0.68. 

We note that vD17 measured the size of the DF44 and DFX1 GC systems finding that  R$_{GC}$/R$_e$ = 1.5 was a good fit to the data. A ratio of 1.5 was found to be suitable for the GC systems of Virgo dwarf galaxies and was adopted for the Coma UDG analysis by L18.
A third study of 54 UDGs,  also using HST,  in the Coma cluster by \cite{2018MNRAS.475.4235A} 
found a mean value of R$_{GC}$/R$_e$ $\sim$2. Thus, on average, GC systems of Coma UDGs appear to be slightly larger than their host galaxy's effective radius. 


This basic analysis does not give us a definitive answer on which study is more correct but it does highlight that GC system size is the reason for the GC count differences between S22 and other literature values. {\it Any future re-analysis of these data should pay particular attention to the size of the GC system and include this in the final uncertainty for their total GC count. }

At a distance of 100 Mpc to the Coma cluster, the HST imaging only indicates candidates for GCs and it is very unlikely that radial velocities will be obtained for any of them in the near future. 
We have visually inspected the deep HST images of DF44 and DFX1 (see figure 1 of vD17), 
for which the faintest compact sources are around the expected turnover magnitude of the GC luminosity function. From this visual inspection we estimate total GC counts similar to those published by vD17. 
For the lower GC counts claimed by S22 to be correct, many of the compact bright sources 
seen in projection on the main body of each galaxy would need to be contaminants. If they are foreground stars or unresolved background galaxies, they would need to match well the colours and magnitudes expected of GCs at 100 Mpc, which is unlikely. Unbound GCs associated with the Coma cluster itself (intracluster GCs) may be a possible source of contaminants and an explanation for the different GC number estimates. 
For the purposes of this work we adopt the GC counts of vD17 and L18 (as noted in section 4), and do not use those of S22.

Another UDG for which there have been different estimates in the literature for its number of  GCs is NGC5846\_UDG1 in the NGC 5846 group. It was 
identified as a UDG with a system of GCs in the deep ground-based imaging of 
\cite{2019A&A...626A..66F}.
A crude estimate of the total number of GCs from this imaging was 45. At $\sim$25 Mpc, this UDG is sufficiently nearby for GCs to be partially resolved with HST. From HST imaging, 
\cite{2021ApJ...923....9M}
derived a total of 26$\pm$6 GCs (which they refer to as MATLAS-2019). Subsequently, with deeper HST imaging, 
\cite{2022ApJ...927L..28D}
derived 54$\pm$9 GCs. In Fig. 4 we show a colour image of NGC5846\_UDG1 from the HST imaging of 
\cite{2022ApJ...927L..28D}
and its exceptionally rich system of GCs. 
They estimated that only 1 background contaminant is contained within their list of detected GCs brighter than the turnover. Thus under the assumption of a universal GCLF the total number of GCs can be found with a high degree of confidence. 
\cite{2022ApJ...927L..28D} had the benefit of deeper (two orbits vs one) and higher resolution (0.04"/pix vs 0.05"/pix) data. Additionally, 
\cite{2021ApJ...923....9M} 
used a narrower colour selection, a higher detection threshold (6.7$\sigma$) and measured a slightly smaller GC system extent (i.e. 0.7 vs 0.8 R$_{GC}$/R$_e$). All of these factors may contribute to the differences in total GC count.

Fortunately, NGC5846\_UDG1 is sufficiently close to obtain spectra of its GC candidates. 
\cite{2020A&A...640A.106M}
confirmed 11 individual GCs, two GCs those spectra were combined together using VLT/MUSE data. A preliminary analysis of Keck/KCWI data has confirmed the 
GCs observed by 
\cite{2020A&A...640A.106M}
(except for one not in the KCWI field-of-view) and added another dozen (Haacke, L. 2023, priv. comm.). Thus the majority of GCs brighter than the GCLF turnover are now confirmed. So irrespective of contamination at the fainter magnitudes, the assumption of a symmetric universal GCLF implies a rich GC system on the order of 50 GCs. Thus here we the GC number of \cite{2022ApJ...927L..28D}, i.e. 54 $\pm$ 9.

\section{Globular Cluster Counts for Individual UDGs}

Next we focus on some individual UDGs with claims in the literature to have over 20 GCs {\it and} to have a velocity dispersion measurement -- either from the 
host galaxy stars or the GC system itself.
These requirements leave us with only three UDGs in the Coma cluster, i.e. DF44 and DFX1 discussed above, and Y358. Recently, 
\cite{2023ApJ...951...77T} (hereafter T23) 
studied the kinematics of the GC systems of 9 UDGs in the Virgo cluster (and 1 brighter NUDGE; VCC1448). Three of their UDGs and VCC1448 host more than 20 GCs according to 
\cite{2020ApJ...899...69L}. 
\cite{2022MNRAS.510..946G}
have measured the stellar velocity dispersions for four  Perseus cluster UDGs -- two of which (R84 and S74) have estimates for their GC counts. We also include NGC5846\_UDG1 which hosts a rich GC system inferred from deep HST imaging, and has both GC system and stellar velocity dispersions available. Other than UDG1 located in the nearby NGC 5846 group, all of our sample are located in dense cluster environments. 

A summary of each UDG's key properties is given in Table 2. This includes our choice of GC number from the literature. From this we also list the GC system mass relative to the galaxy stellar mass (assuming a mean GC mass of 2 $\times$ 10$^5$ M$_{\odot}$) and 
the inferred halo mass based on GC number using the GNHR of BF20. We include velocity dispersion from stars and/or the GC system and galaxy half-light size. We list the dynamical mass following the method of Wolf et al. (2010) which assumes a pressure-supported system in dynamical equilibrium. Like most UDGs, our sample galaxies are nearly round on the sky with mean b/a $\sim$ 0.8. Total halo masses assuming cusp or core profiles, extrapolated from the dynamical mass are also listed in Table 2 (see Section 6 for details).
Details for the individual galaxies are given below.

\subsection{DF44}

As noted in the Introduction, DF44, in the Coma cluster, is the only UDG to have published radially extended kinematics and the only one to have an independent dynamical measure of its total halo mass. From deep exposures on the Keck II telescope using KCWI, 
\cite{2019ApJ...880...91V}
measured a velocity dispersion profile out to $\sim$ 5~kpc with no sign of bulk rotation. This data was well fit by a cored dark matter halo which indicated isotropic orbits giving a total halo mass of log M$_h$ = 11.2 $\pm$ 0.6. Isotropic orbits were preferred from the shape of the absorption lines. 
A cuspy NFW fit to the data required strongly tangential orbits and gave a lower total halo mass of log M$_h$ = 10.6$^{+0.4}_{-0.3}$. 
The published studies of GC counts for DF44 range from  20$^{+6}_{-5}$ from S22 to 76 $\pm$ 18 from vD17. 
However, as discussed above, we have elected to adopt the vD17 GC number.  

\subsection{DFX1}

DFX1 is located in the Coma cluster.
As noted in Table 1, vD17 estimated DFX1 to host 63 $\pm$ 17 GCs, while S22 suggested 17$^{+5}_{-6}$ GCs. Again, given the general agreement between vD17 and L18, we adopt the vD17 GC count.

\subsection{Y358}

Yagi358 (or Y358) is a Coma cluster UDG from the 
\cite{2016ApJS..225...11Y}
catalog. Taking the average of the vD17 and L18 GC counts, we list 28 $\pm$ 5.3 GCs in Table 2 for Y358. 

\subsection{VLSB-B} 

Previous selection of GC candidates by 
\cite{2020ApJ...899...69L}
from imaging indicated a total GC system of 26.1 $\pm$ 9.9. 
Recently, T23
obtained Keck/DEIMOS spectra for 14 GCs 
and derived a GC system velocity dispersion of 45$^{+14}_{-10}$ km s$^{-1}$. 
This Virgo cluster galaxy has a recession velocity of +40 km s $^{-1}$ so foreground star interlopers remain a possibility.  A stellar velocity dispersion is not available. 
Following T23 we use their updated distance of 12.7 Mpc to derive an effective radius of R$_e$ = 1.7 kpc after circularisation using $b/a=0.83$~\citep{2018ApJ...856L..31T}.



\subsection{VCC1287} 

Multi-filter, deep ground-based imaging of VCC1287 in the Virgo cluster has been carried out by 
\cite{2016ApJ...819L..20B} and \cite{2020ApJ...899...69L}. 
The former estimated 22 $\pm$ 8 GCs, while the latter suggested 27.6 $\pm$ 11.1 GCs. These estimates are consistent within the quoted uncertainties (both are around 40\%), here we take an average and list a GC count of 25 $\pm$ 5 in Table 2. 

Beasley et al. also obtained spectroscopy of the brighter GCs using the 10m 
Gran Telescopio de Canarias (GTC), confirming 6 to be GCs associated with the galaxy and the galaxy nucleus. Thus a significant fraction of the  brightest GC candidates have been confirmed. They measured a GC system velocity dispersion of 33$^{+16}_{-10}$ km s$^{-1}$. Using the same GCs and applying their own calculation method, T23 derived $\sigma_{GC}$ = 39$^{+20}_{-12}$ km s$^{-1}$. Here we adopt the 
\cite{2016ApJ...819L..20B}
value given its lower formal errors. With only 8 GCs, and large associated uncertainties, a derived dynamical mass should be considered quite uncertain. A stellar velocity dispersion is also available from 
\cite{2020MNRAS.495.2582G}
who measured 19$\pm$6 km s$^{-1}$, significantly lower than the GC system value.

\subsection{VCC615} 

Of the total GC system of 30.3 $\pm$ 9.6 estimated by 
\cite{2020ApJ...899...69L}, 
T23 measured the radial velocities for 8 GCs. From these  they calculated a GC system velocity 
dispersion $\sigma$ = 36$^{+22}_{-18}$ km s$^{-1}$. With only 8 GCs, and large associated uncertainties, a derived dynamical mass should be considered quite uncertain. A stellar velocity dispersion for VCC615 is not available. 
We follow T23 and use a distance of 17.7 Mpc which places VCC615 at the far side of the Virgo cluster and take their effective radius of R$_e$ = 2.0 kpc. 


\subsection{VCC1448}

At $\sim$1 mag brighter than the standard UDG definition and with 
R$_e$ = 3.5 kpc (T23), 
VCC1448 is an example of a NUDGE galaxy, located in the Virgo cluster. 
Its stellar mass of M$_{\ast}$ = 2.6 $\times$ 10$^9$ M$_{\odot}$ places it in the low mass galaxy regime yet it has a remarkably rich GC system with 
\cite{2020ApJ...899...69L}
estimating 99.3 $\pm$ 17.6 GCs. Thus it appears to host around half the number of GCs associated with the Milky Way. 
T23 measured a GC system velocity dispersion from 9 GCs to be 48$^{+16}_{-11}$ km s$^{-1}$. However, Gannon, J. (2023, priv. comm.) measure a 
stellar velocity dispersion. 
They also find evidence of bulk rotation in the galaxy at the $\sim$20 km s$^{-1}$ level. 
In Table 2 we calculate 
the dynamical mass using 
the combined rotation and dispersion velocities.

\subsection{PUDG-R84 and PUDG-S74}


\cite{2022MNRAS.510..946G}
measured the 
stellar velocity dispersions for four 
Perseus UDGs. Two of them have 
estimates of their GC system in excess of 20. 
For PUDG-R84, HST imaging is available in two filters. The total number of GCs found by Janssens, S. (2023, priv. comm.) is 42$\pm$6.
PUDG-S74 is not covered by the HST imaging and here we use a visual estimate by 
\cite{2022MNRAS.510..946G} based on ground-based imaging to be around 30 GCs.

\begin{figure}
	\includegraphics[width=\linewidth]{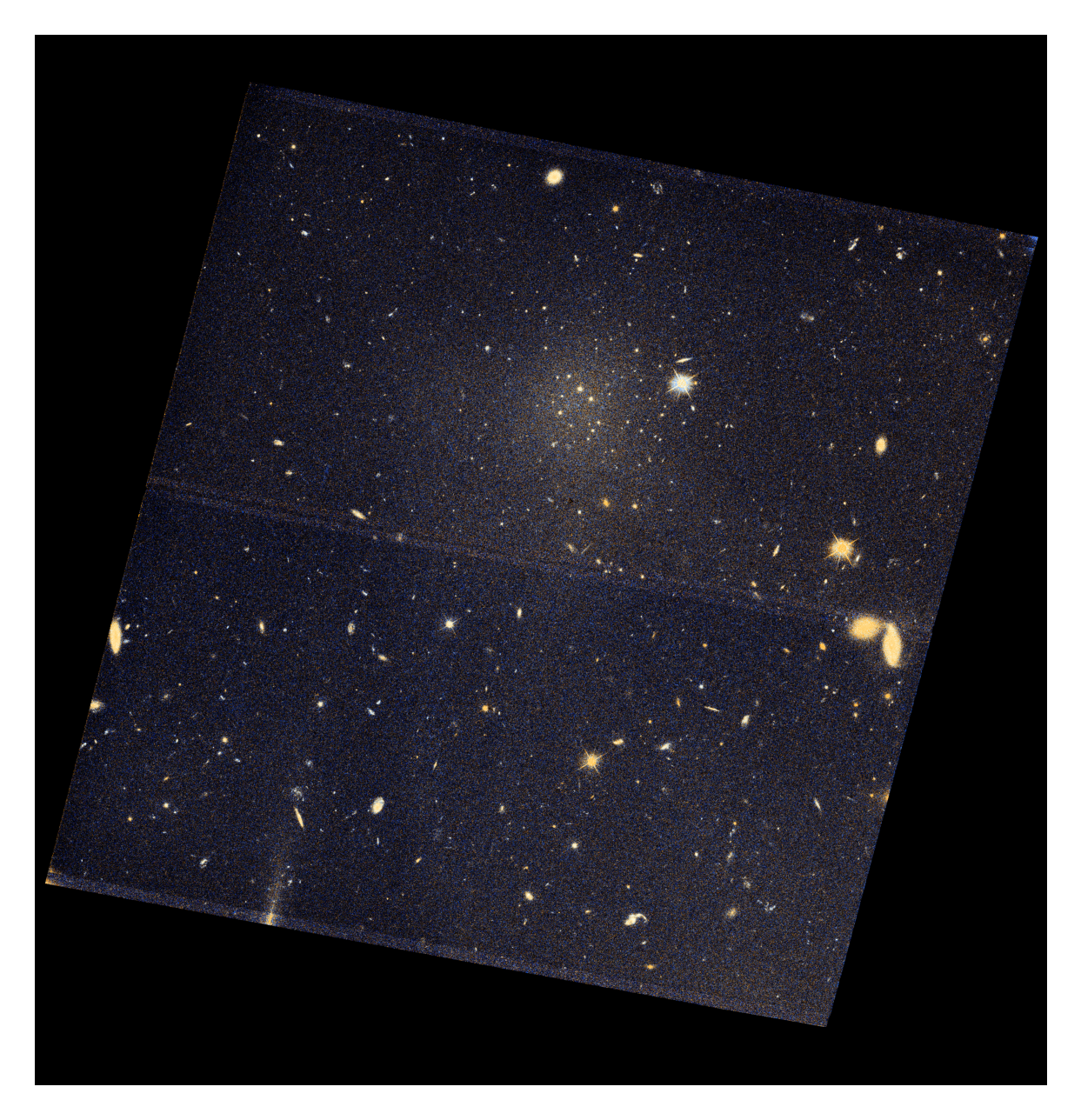}
	\caption{Colour $(g-V)$ HST/WFC3 image of NGC5846\_UDG1. North is up and East is left. 
 The $\sim$2.7 $\times$ 2.7 sq. arcmin image shows a rich system of GCs around the UDG compared to the outer regions of the image. The total GC count estimated by 
 \protect\cite{2022ApJ...927L..28D}
 is 54 $\pm$ 9 GCs.  The majority of the brightest 27 GCs have been confirmed by radial velocities (Haacke, L. 2023, priv. comm.)  
 }
 \end{figure}
 
\subsection{NGC5846\_UDG1}

As mentioned above, we adopt the GC count from 
\cite{2022ApJ...927L..28D} and list 54 $\pm$ 9 
for NGC5846\_UDG1 in  Table 2. 
This galaxy benefits from both GC system and stellar velocity dispersion measurements. 
\cite{2021MNRAS.500.1279F} 
measured a stellar velocity dispersion of 17 $\pm$ 2 km s$^{-1}$. From 11 GCs 
\cite{2020A&A...640A.106M}
measured 9.4$^{+7.0}_{-5.4}$ km s$^{-1}$ however more recent observational work (Haacke, L. 2023, priv. comm.), using around 20 GCs, suggests a higher value more in line with the stellar velocity dispersion. 

\begin{table*}
  \centering
  \begin{tabular}{l|c|c|c|c|c|c|c|c|c|c|c}
  \hline
      UDG & N$_{GC}$ & M$_{h}$(GC) & M$_{\ast}$ & M$_{GC}$/M$_{\ast}$ & R$_e$ & $\sigma_{\ast}$ & $\sigma_{GC}$ & Ref. & M$_{\rm dyn}$ & M$_{h}$(cusp) & M$_{h}$(core) \\
      & & (10$^{10}$ M$_{\odot}$) & (10$^{8}$ M$_{\odot}$) & (\%) & (kpc) & (km/s) & (km/s) & & (10$^{9}$ M$_{\odot}$) &(10$^{10}$ M$_{\odot}$) & (10$^{10}$ M$_{\odot}$)\\
      \hline 

DF44 & 76$\pm$18 & 38 & 3.0 & 5.0 & 3.9 &  33$\pm$3 & -- & vD19 & 3.95 & 4.46 & 77.8\\
DFX1 & 63$\pm$17 & 32 & 3.4 & 3.8 & 2.8 & 30$\pm$7 & -- & vD17 & 2.34 & 4.01 & 100\\
Y358 & 28$\pm$5.3 & 14 & 1.4 & 4.0 & 2.1 & 19$\pm$3 & -- & G23 & 0.71 & 0.66 & 9.39\\
VLSB-B & 26.1$\pm$9.9 & 13 & 0.22 & 23 & 1.7 & -- & 45$^{+14}_{-10}$ & T23 & 3.20 & 223 & 30053\\
VCC1287 & 25$\pm$5 & 13 & 3.9 & 1.2 & 3.3 & 19$\pm$6 & 33$^{+16}_{-10}$ & G20, B16 & 1.12 & 0.56 & 4.67\\
VCC615 & 30.3$\pm$9.6 & 15 & 0.73 & 8.4 & 2.3 & -- & 36$^{+22}_{-18}$ & T23 & 2.77 & 15.5 & 949\\
VCC1448$^{\dagger}$ & 99.3$\pm$17.6 & 50 & 26 & 0.8 & 3.15 & 24$\pm$1.4 & 38$^{+9}_{-6}$ & T23, G24  & 1.92 & 1.76 & 25.03\\
PUDG-R84 & 42$\pm$6 & 21 & 2.2 & 3.8 & 2.0 & 19$\pm$3 & -- & G22 & 0.67 & 0.69 & 10.43\\
PUDG-S74 & $\sim$30 & 15 & 7.9 & 0.8 & 3.5 & 22$\pm$2 & -- & G22 & 1.58 & 0.93 & 8.94\\
NGC5846$\_$UDG1 & 54$\pm$9 & 27 & 1.1 & 9.8 & 2.1 & 17$\pm$2 & 9.4$^{+7.0}_{-5.4}$ & F21, M20 & 0.56 & 0.42 & 4.89\\
     \hline
     
  \end{tabular}
  \caption{Literature data for UDGs with both N$_{GC}$ $>$ 20 and a measured velocity dispersion. Columns list UDG name, number of GCs (see text), halo mass inferred from GC number, stellar mass (T23 for Virgo galaxies or G23 for others), effective radius (from G23 or T23 for VLSB-B, VCC615 and VCC1448), stellar velocity dispersion, GC system velocity dispersion, and velocity dispersion reference. The dynamical mass and inferred halo masses use the stellar velocity dispersion when available. 
   Last two columns list halo mass under the assumption of a cusp and core mass profile. The GC systems of VLSB-B and VCC615 do not appear to be in dynamical equilibrium. 
$\dagger$: VCC1448 does not meet the UDG definition. For VCC1448 we calculate M$_{dyn}$ by combining its stellar velocity dispersion and stellar rotation in quadrature from Gannon, J. (2023, priv. comm.). See  https://github.com/gannonjs/Published$\_$Data/tree/main/
UDG$\_$Spectroscopic$\_$Data for further basic UDG data.
}
\end{table*}

\section{Do UDGs with more GCs have higher dynamical masses?}

From the above discussion we conclude that at least {\it some} UDGs do indeed host rich GC systems of over 20 GCs.
Nearby UDGs (e.g. those in Virgo cluster or the NGC 5846 group) have the benefit of HST imaging which can partially resolve their GCs and of spectroscopy which can give radial velocities.  The rich GC systems listed in Table 2 imply very high ratios of GC system mass to host galaxy stellar mass. This ratio ranges from 0.8\% to 23\% for VLSB-B with its 26.1 GCs (we assume a mean GC mass of 2 $\times$ 10$^5$ M$_{\odot}$). On average the 9 UDGs (we exclude VCC1448) have 6\% of the galaxy stellar mass in the mass of GCs. This is some 2--4 times the average ratio seen in classical dwarf galaxies with the same stellar mass 
(\citealt{2020ApJ...899...69L}; \citealt{2020MNRAS.492.4874F}). 
We now turn to the question of whether GC-rich UDGs also have higher host galaxy dynamical masses (which might suggest higher total halo masses as well).

\cite{2013ApJ...772...82H}
found that the number of GCs associated with a galaxy shows a  strong correlation with the enclosed dynamical mass within the half-light radius. Their sample consisted of $\sim$170 normal galaxies, dominated by bulges and spheroids (i.e. pressure supported systems). Dynamical masses were calculated using the formula of 
\cite{2010MNRAS.406.1220W}
from measurements of the bulge/spheroid $\sigma_e$ and R$_e$. The relationship between N$_{GC}$ and M$_{dyn}$ shows
distinctly different slopes at around a dynamical mass of log M$_{dyn}$ $\sim$ 11 M$_{\odot}$ (see their figure 9). 
We have fit their sample for normal galaxies for masses lower than this inflection point, finding:\\

\noindent
log N$_{GC}$ = 0.64~log M$_{dyn}$ -- 4.59\\


Turning now to our literature sample of GC-rich UDGs, we adopt the same 
\cite{2010MNRAS.406.1220W}
formula to calculate their dynamical masses and we include them in Table 2. 
We use stellar velocity dispersions when available, if not, we use GC system ones. The stellar velocity dispersions are not subject to bias by interlopers (foreground stars and/or intra-cluster GCs) nor small number statistics. 
The Wolf et al. formula assumes sphericity and stationarity, along with a pressure-supported system with no rotation and quasi-constant velocity dispersion profiles 
(for a full discussion of the assumptions and limitations see \citealt{2010MNRAS.406.1220W}). 
When we use GC system velocity dispersions we assume that they are equivalent dynamical tracers to the galaxy stars. Following the approach of most GC studies, and 
\cite{2013ApJ...772...82H},
we use the host galaxy half-light radius as the physical size scale in the formula. 
In their dynamical analysis, 
\cite{2023ApJ...951...77T} (hereafter T23) 
used the GC system half-number radius. 

Comparing GC system and stellar velocity dispersions for a same galaxies with both measurements,  
\cite{2021MNRAS.500.1279F} 
found some scatter but little evidence of a systematic bias. 
The simulations of 
\cite{2021MNRAS.502.1661D}
also support this conclusion. The main caveat is perhaps when the dynamical mass is based on radial velocities of fewer than 10 GCs (as is often the case in the T23 study). In this case,  
\cite{2021MNRAS.502.1661D}
showed that the dynamical mass can be either over, or under, estimated compared to the true value depending on the method used. We note that T23 attempted to correct for this bias in their dynamical analysis, nevertheless small number statistics remain a concern. 

In Fig. 5 we show the number of GCs in our GC-rich UDGs versus enclosed dynamical mass and the fit to the low mass normal galaxies from the sample of 
\cite{2013ApJ...772...82H}.
The typical uncertainty of the dynamical is around 0.1 dex. 
The right hand axis shows the halo mass predicted 
from the GNHR of BF20.
{\it We find a general trend for UDGs with more GCs to have higher dynamical masses.}
Given that UDGs are known to be DM dominated in their inner regions (\citealt{2018ApJ...856L..31T}; 
\citealt{2021MNRAS.502.3144G}), 
the trend of increasing dynamical mass with increasing GC count suggests that a similar trend holds with total halo mass. 
The same conclusion was reached by 
\cite{2022MNRAS.510..946G}
for a sample that included some GC-poor UDGs.  

Although the UDGs lie within the scatter of normal galaxies 
they are systematically offset from normal galaxies. 
While the dynamical masses of UDGs, within the half-light radius, are DM dominated, those of normal galaxies may be baryon dominated. This can be seen in 
\cite{2013ApJ...772...82H}
where the dynamical mass has a near linear correlation with K-band magnitude (their figure 4), a good proxy of stellar mass. Thus in Fig. 5  
UDGs can be interpreted as having higher GC counts, by a factor of about two, compared to normal galaxies of the same dynamical mass. A similar conclusion is reached when comparing GC numbers, or GC system mass, in UDGs to classical dwarf galaxies at a given stellar mass (e.g. 
\citealt{2020MNRAS.492.4874F}). 
This indicates an enhanced efficiency of 
GC formation in UDGs compared to normal galaxies. 
This conclusion does not appear to be solely driven by the large half-light radii  of UDGs. For example, placing our GC-rich UDGs on the right hand panel of figure 7
(N$_{GC}$ vs velocity dispersion) from 
\cite{2013ApJ...772...82H} indicates that, again, our GC-rich UDGs host more GCs at a given galaxy velocity dispersion compared to normal galaxies.

\begin{figure}
	\includegraphics[width=1.05\linewidth]{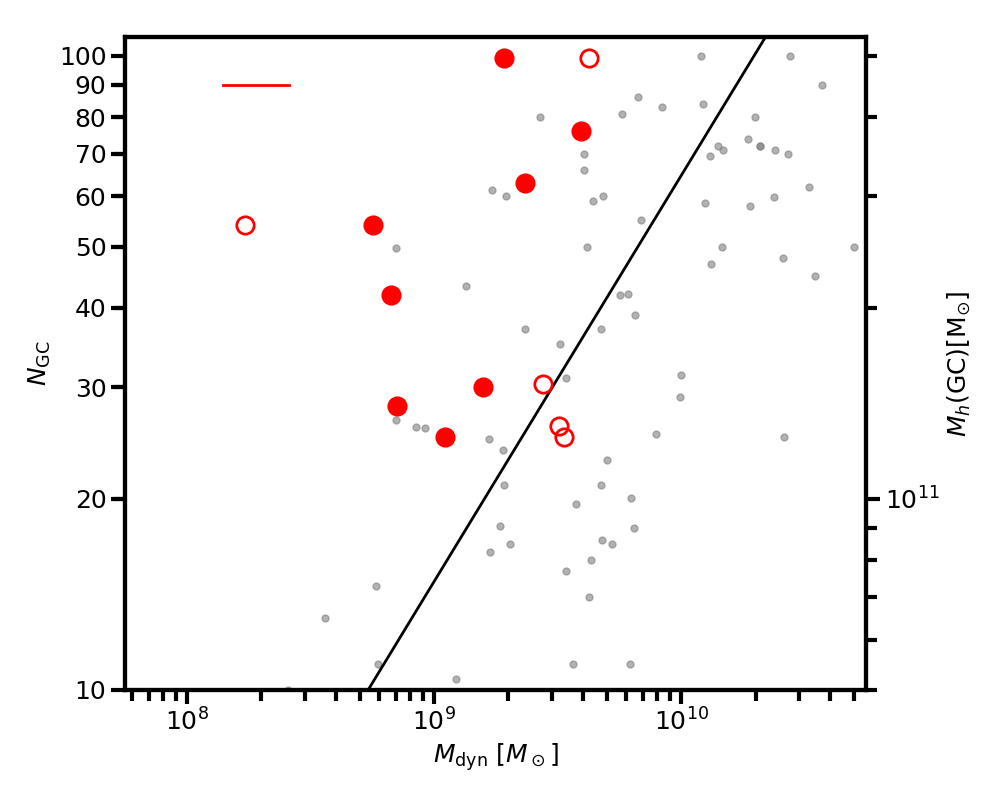}
	\caption{Number of GCs as a function of dynamical mass within the half-light radius. The right hand side axis shows the total halo mass (assuming the relation of BF20). Red symbols show various UDGs with measured dynamical masses and $>$ 20 GCs (filled circles for stellar and open circles for GC system velocity dispersions respectively). Three galaxies are shown twice on this plot, i.e. they have both stellar and GC-based dynamical masses. The typical uncertainty on the stellar-based dynamical masses is indicated in the top left. 
  We include VCC1448 (N$_{GC}$ = 99) which is UDG-like in size but  $\sim$1~mag brighter than the UDG definition. Normal galaxies from 
   \protect\cite{2013ApJ...772...82H}
   are shown by small grey symbols. The 
  solid line shows the fit to the low mass normal galaxies. There is a trend for UDGs with more GCs to have higher dynamical masses. UDGs also have more GCs at a given dynamical mass, than normal galaxies.
  }
\end{figure}

\section{Deriving Halo Masses from Measured Dynamical Masses}





Our finding that UDGs with more GCs also have higher dynamical masses suggests that they also have higher total halo masses. Next we use the measured dynamical masses to infer the total halo mass by assuming 
a halo mass profile. 
In general, this mass profile can either be cuspy like the standard NFW profile 
\citealt{1997ApJ...490..493N}, 
or cored due to baryonic heating (e.g. 
\citealt{2016MNRAS.459.2573R}; \citealt{2017MNRAS.466L...1D}).
For a recent review of cusps and cores see 
\cite{2022arXiv220914151D}.


Here we generate cuspy NFW halo and core-NFW halo profiles following 
\cite{2016MNRAS.459.2573R}.
The NFW profiles are generated using Appendix A of 
\cite{2014MNRAS.441.2986D}.
The core-NFW halo profiles are created as a modification of the basic NFW profile as described in 
\cite{2016MNRAS.459.2573R}.
To set the core radius we follow 
\cite{2017MNRAS.467.2019R}
who showed that the ratio of core radius to the half-light radius increases with halo mass (their figure B1) rising to be close to the maximum allowed by SN energy arguments. Here we assume this maximal core formation and set each core radius to be 2.75 times the observed half-light radius of the UDG. 
For halo concentrations (c) we follow the standard prescription that it varies inversely (but weakly) with total halo mass according to 
\cite{2014MNRAS.441.3359D}. We note that cored profiles have lower effective concentrations than cuspy profiles.

The mass profiles fit to the measured dynamical mass, within the half-light radius, for the GC-rich UDGs are shown in Fig. 6. The plot shows the cumulative mass profile assuming cusp and core halo profiles. We preference stellar over GC system velocity dispersions to calculate masses in Table 2 but show both in Fig. 6. 
The uncertainty on the derived halo mass is dominated by the extrapolation from the dynamical mass within the half-light radius to the virial radius, so that  
a small difference in the half-light radius and/or velocity dispersion can lead to large differences in the resulting halo mass (e.g. 
\citealt{2005MNRAS.363..705M}). 
The resulting halo masses, for both cusp and core profiles, are listed in Table 2. The uncertainty on halo masses extrapolated from dynamical masses should be considered to be an order of magnitude. 

For two UDGs (VCC615 and VLSB-B), the halo mass based on their GCs from T23 results in unrealistically high halo masses (see Table 2). T23 appealed to high concentrations to explain their high halo masses, which in some cases deviated by more than 3$\sigma$ from the c--M$_{h}$
relation of \cite{2014MNRAS.441.3359D}.  
For VCC1287 and VCC1448 the stellar velocity dispersion, and hence the inferred dynamical mass, is measured to be much lower than using the GC system values from T23. 

The recent simulations of 
\cite{2023arXiv230903260D}
focused on the observations T23. While they concluded that GCs should trace the underlying DM halo well, they also found the presence of intracluster GCs could explain some of the high velocity dispersions reported by T23 and, that when fewer  than 10 GCs were observed the velocity dispersion can be highly unreliable. They didn't explore the effects of the spatial sampling of T23, which tends to select relatively more GCs in the galaxy outer regions due to the Keck/DEIMOS field-of-view. This latter effect may also increase the number of interloper GCs in the T23 sample.  
We suspect that the velocity dispersions for many of the T23 GC systems are overestimated and/or not in dynamical equilibrium.


It is currently unknown whether UDGs occupy cuspy or cored halo mass profiles.
Around 1/3 of UDGs are known to reveal nuclei 
\citet{2021A&A...654A.105M}.
If these nuclei are the result of infalling and merged GCs then this might suggest cuspy mass profiles are present (which facilitate a shorter dynamical friction timescale). 
We also note that 
\cite{2023arXiv230702530D}
were able to reproduce galaxies with high GC system mass to stellar mass ratios approaching 10\% (as seen in some GC-rich UDGs) when dynamical friction was turned off. This might indicate that such high ratios could be a signature of a cored DM halo in which the timescale for GC merging via dynamical friction is extremely long. 
In the Local Group, the Fornax dwarf galaxy (a NUDGE) is generally thought to be an example of a galaxy with a cored DM profile from its lack of a nucleus and its stellar kinematics (see 
\citealt{2016MNRAS.459.2573R}
and references therein), although initial conditions 
\citep{2021PhRvD.104d3021B}
and priors 
contribute some uncertainty to this interpretation.
We also note the detailed study of a field UDG by \citet{2021ApJ...919L...1B}. Using its HI kinematics, they derived a shallower slope than a canonical NFW cusp and with a halo concentration that matched the expectations of \cite{2014MNRAS.441.3359D}
, whereas an NFW cusp solution was some 5$\sigma$ away from the \cite{2014MNRAS.441.3359D} 
relation.

There are some good theoretical reasons for expecting UDGs to occupy cored halos. 
Dwarfs and low surface brightness galaxies up to log M$_{\ast}$ = 9 (i.e. the regime of UDGs) are generally expected to reveal cored profiles 
(\citealt{2011AJ....142...24O}; \citealt{2012MNRAS.422.1231G}).
More recent hydrodynamical simulations also predict mass profiles to be dominated by cores at the stellar masses of UDGs  (\citealt{2016MNRAS.456.3542T}).
Indeed, the creation of cores, via SN feedback, is a key element in several simulations of UDGs, e.g. isolated UDGs in the NIHAO models of 
\cite{2017MNRAS.466L...1D} and group/cluster UDGs susceptible to tidal heating in the models of 
\cite{2019MNRAS.485..796M}.




\section{The Stellar Mass -- Halo Mass Relation (SMHR) and UDGs}

We now investigate the location of our GC-rich UDGs relative to the standard SMHR for normal galaxies. In this work we have calculated total halo masses from three different methods: \\

\noindent
$\bf 1)$ GC number assuming the GNHR\\
$\bf 2)$ dynamical mass assuming a cuspy halo profile\\
$\bf 3)$ dynamical mass assuming a cored halo profile\\

\begin{figure}
	\includegraphics[width=1.0\linewidth]{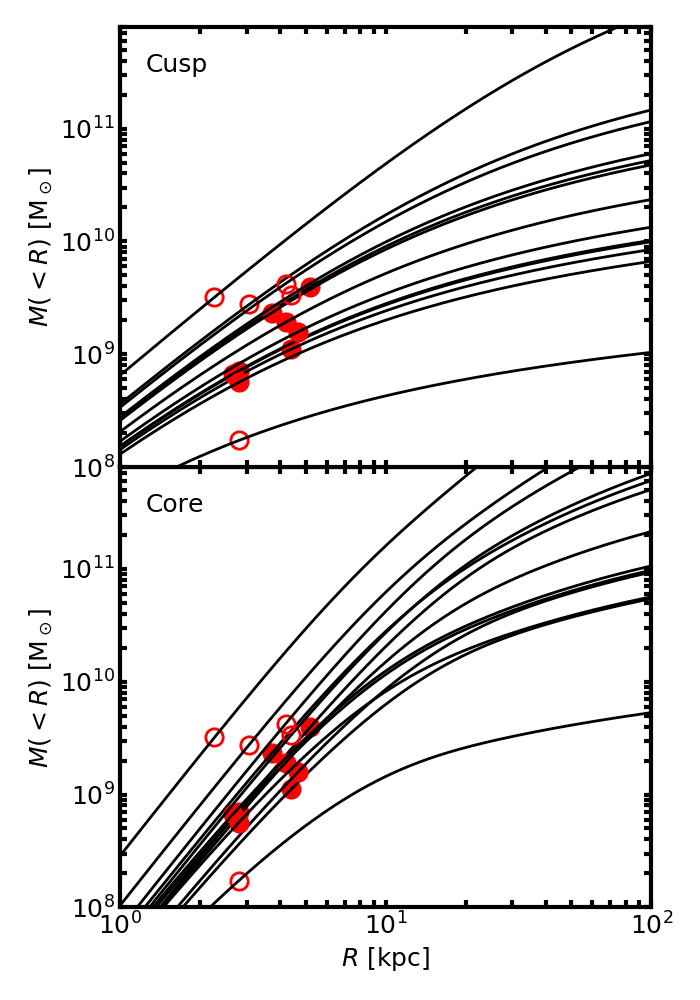}
	\caption{Cumulative mass profiles for GC-rich UDGs assuming cusp ({\it upper}) and core ({\it lower}) halos.
  Red symbols show the measured dynamical masses for our sample of GC-rich UDGs (filled circles using stellar, and open circles using GC system, velocity dispersions respectively). 
    Solid lines show the fit to the dynamical masses assuming a cusp or core halo profile. The two highest mass profiles in each panel correspond to VLSB-B and VCC615, which may not be in dynamical equilibrium.  
  }
\end{figure}


\begin{figure}
	\includegraphics[width=1.05\linewidth]{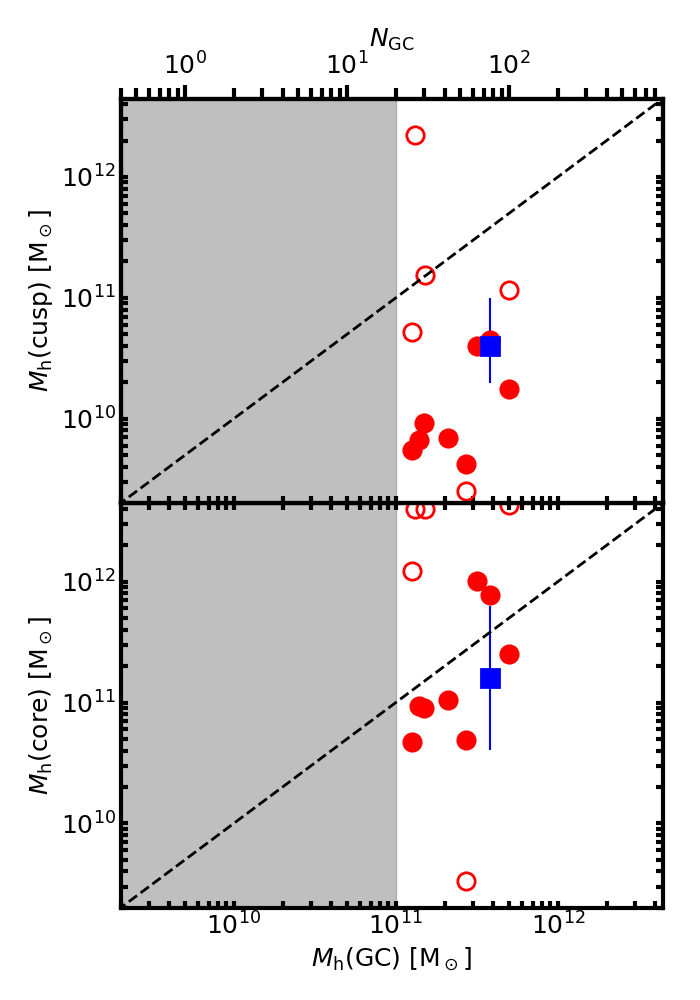}
	\caption{Cuspy and cored halo masses versus halo mass from the GNHR for GC-rich UDGs. {\it Upper panel} halo masses derived from the assumption of a 
 cuspy mass profile and {\it lower panel} halo masses from the assumption of a core mass profile.  The upper axis gives the number of GCs corresponding to the halo mass using the GNHR of BF20. The shaded region is excluded from our selection, i.e. our sample has N$_{GC}$ $>$ 20. Filled and open symbols are based on stellar and GC system velocity dispersions respectively. Masses above log M$_{h}$ = 12.5 are displayed at the extreme of the y axis. The blue squares represent the cusp and core mass solutions for DF44.
 The dashed line is a unity line in each panel. The agreement with halo masses from GC numbers is better for halo masses derived assuming cored profiles. 
}
 \end{figure} 

\begin{figure*}
	\includegraphics[width=1.0\linewidth]{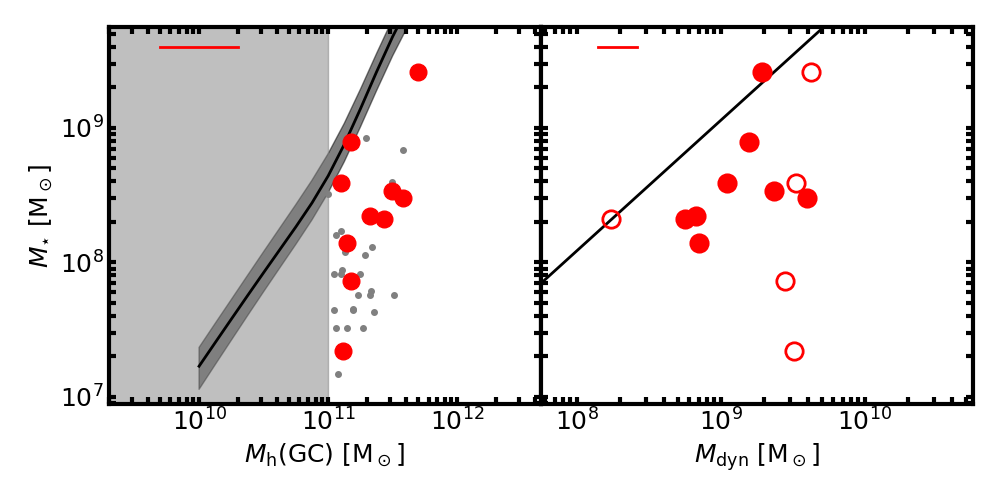}
	\caption{Stellar mass -- halo mass relation (SMHR) and dynamical mass relation.
 Red symbols show UDGs with more than 20 GCs from Table 2 as described in the text (open symbols for GC system based velocity dispersions). The galaxy with the highest stellar mass, VCC1448, is not classified as a UDG. {\it Left} Halo masses derived from GC numbers by applying the scaling relation of BF20. A typical error bar (upper left) indicates the 0.3 dex uncertainty from the scatter in the GNHR. 
 The SMHR for normal galaxies from  
 \protect\cite{2013ApJ...770...57B} is shown as a dark line 
 with 1$\sigma$ scatter.  
 This popular SMHR is based on $\Omega_M$ = 0.27, however $\Omega_M$ = 0.31, favoured in Planck cosmology, would make only a small difference.  The shaded region is excluded from our selection, i.e. our sample has N$_{GC}$ $>$ 20 (M$_h$ $>$ 10$^{11}$ M$_{\odot}$). Small grey symbols show Coma cluster UDGs with more than 20 GCs from the compilation of 
 \protect\cite{2020MNRAS.492.4874F}.
 These UDGs, along with the UDGs from Table 2, deviate on average from the standard SMHR with overly massive halos. 
 {\it Right} Total stellar mass vs measured dynamical mass within the half-light radius. 
 The typical uncertainty for the dynamical mass of the filled symbols is shown in the upper left. 
 The solid line is a fit to the low mass normal galaxies of 
 \protect\cite{2013ApJ...772...82H}. 
 For a given stellar mass, UDGs tend to have higher 
 higher dynamical masses than normal galaxies. 
 }
\end{figure*}

We start in Fig. 7 comparing the halo masses derived from GC number (method $\bf 1$) with those by assuming a cuspy (method $\bf 2$) or cored (method $\bf 3$) mass profile. We take the uncertainty on the halo masses from GC counts as the scatter in the GNHR, i.e. 0.3 dex, as dominates over the measurement uncertainty in GC counts. 
Halo masses assuming core, or cuspy, profiles have large, order of magnitude uncertainties. {\it The figure shows that cored halo masses, particularly those that use stellar velocity dispersion based dynamical masses, are in good agreement with halo masses inferred from GC counts.} 
Cuspy profiles tend to give halo masses an order of magnitude below those predicted from GC counts.  

As mentioned above, a total halo mass measurement from mass modelling of the kinematic profile exists for one UDG, i.e. DF44 
(\citealt{2019ApJ...880...91V}).
From the shape of the absorption lines, isotropic orbits were favoured over tangential ones. 
The fit for a cored mass profile was preferred as it required isotropic orbits, giving a best fit halo mass of log M$_{h}$ = 11.2$^{+0.6}_{-0.6}$. A cuspy NFW mass profile gave a halo mass of 
log M$_{h}$ = 10.6$^{+0.4}_{-0.3}$ but required disfavoured  tangential orbits. We show these two halo masses on the appropriate panels of Fig. 7 with a blue square symbol. 
The favoured core solution is consistent with the halo mass derived from GC numbers. 

A comparison of SMHRs, derived from various methods, has been summarised by 
\cite{2018ARA&A..56..435W}.
While large differences exist at the highest and lowest masses between studies, there is fairly good agreement at log M$_{\ast}$ $\sim 8$, i.e. the masses of UDGs. 
Here we adopt the well-known SMHR from 
\cite{2013ApJ...770...57B}.
Only small variations in the predicted halo mass, at the stellar masses of UDGs, would be found if other SMHRs were used.
We now proceed to examine the location of GC-rich UDGs in comparison to this standard SMHR.

The left panel of Fig. 8 shows the location of our GC-rich UDGs relative to the SMHR of 
\cite{2013ApJ...770...57B} with halo masses derived from GC counts.
Thus we assume the GNHR is valid for GC-rich UDGs (method $\bf 1$). 
{\it The plot shows that most of our sample GC-rich UDGs deviate from the standard SMHR with overly massive halos for their stellar mass}. 

For example, NGC5846\_UDG1 has a 
stellar-to-halo mass ratio of 4.1 $\times$ 10$^{-4}$ compared to 
2.8 $\times$ 10$^{-3}$
if it obeyed the standard SMHR.
The average stellar and halo mass for our sample of GC-rich UDGs (i.e. excluding VCC1448) is 8.48 and 11.32 in log respectively. 
Thus the GC-rich UDGs lie several sigma, on average,  from the standard SMHR using GC-based halo masses. We remind the reader that our UDG sample is selected to have more than 20 GCs, which corresponds to halo masses greater than 10$^{11}$ M$_{\odot}$. This cutoff leads to 
the trend for the lowest stellar mass UDGs to deviate the most from the SMHR.
While this might be a highly-selected sample of UDGs, a larger sample of GC-rich Coma cluster UDGs, which probe to lower stellar masses on average, also indicates overly massive halos relative to the standard SMHR
\citep{2020MNRAS.492.4874F}.
Thus finding overly massive halos in the best studied GC-rich UDGs (Table 2) is not a selection effect driven by their higher average masses, sizes or surface brightnesses.

In the right panel of Fig. 8 we show dynamical mass within the half-light radius. The GC-rich UDGs have typical dynamical masses of 10$^9$ M$_{\odot}$ compared to their {\it total} stellar masses of 10$^8$ M$_{\odot}$, indicating they are strongly dark matter dominated within their half-light regions. 
The typical uncertainty on the dynamical mass derived from stellar velocity dispersions (filled symbols) is $\sim$30\%, but can be much higher for the GC system based velocity dispersions (open symbols).
The plot also includes a fit to the low mass normal galaxies from 
\cite{2013ApJ...772...82H}.
This panel shows that GC-rich UDGs have systematically higher dynamical masses, within their half-light radii, compared to normal galaxies of a similar stellar mass. In other words, the UDGs are significantly more dark matter dominated within their half-light radii than normal galaxies. This trend, relative to normal galaxies, resembles that for halo mass based on GC number.

In summary, Fig. 8 suggests that GC-rich UDGs reveal a strong tendency for overly massive halos, relative to the SMHR if they obey the GNHR. This is supported by the finding of higher than average dynamical masses for a given stellar mass compared to normal galaxies. There are theoretical reasons to expect UDGs to occupy cored halos, and halo masses inferred from cored mass profiles agree better than cuspy halos with halo masses inferred from GC numbers (see Fig. 7).
If GC-rich UDGs were to occupy cuspy mass halos then they would 
lie in extremely under massive halos relative to the SMHR, {\it and} they do not obey the GNHR (see also Fig. 7). 
We consider cuspy halos a less likely option for GC-rich UDGs but it can not be ruled out with the available data.

\section{Discussion}

At a given stellar mass, UDGs host more GCs, on average, than normal galaxies. They also reveal higher dynamical masses than normal galaxies, indicating they are more DM dominated within their half-light radii.
As noted in the Introduction, GC-rich UDGs either obey the GNHR or the SMHR, but they can not obey both scaling relations simultaneously, at their measured stellar masses. Therefore, they must be outliers relative to one of these two fundamental scaling relations which hold for normal galaxies over several orders of magnitude. 

For GC-rich UDGs, we have found that halo masses inferred from the GNHR are in reasonable agreement with halo masses derived by assuming cored halo mass profiles (see Fig. 7). 
(The agreement assuming cuspy profiles was very poor.) 
This suggests that GC-rich UDGs do not obey the SMHR, having   overly massive halos for their stellar mass (see Fig. 8).  
In a recent statistical study of $\sim$500 UDGs, 
\cite{2023ApJS..267...27Z} 
reached a similar conclusion finding UDGs have higher M$_{h}$/M$_{\ast}$ ratios than normal galaxies. 
We also find a related trend for higher dynamical masses at a given stellar mass (see Fig. 8).  
{\it We favour the conclusion that GC-rich UDGs obey the GNHR and deviate from the SMHR with overly massive cored halos.}

While there are some good theoretical arguments in favour of UDGs occupying cored halos, we can not rule out cuspy halos. If GC-rich UDGs indeed occupied cuspy halos, then our results indicate that GC-rich UDGs lie in extremely low mass halos,  neither following the SMHR nor the GNHR. We consider this possibility less likely. 

Given the inverse correlation of halo mass and concentration, one could appeal to higher concentrations to bring halo masses down to follow the standard SMHR. This is the approach of T23 who argued for c $>$ 20 to bring their halo masses (derived assuming cuspy mass profiles) down to $\sim$10$^{12}$ M$_{\odot}$. Although in their study, this still implies UDGs have overly massive halos and do not follow the SMHR. It is unclear whether low stellar density galaxies, like UDGs, would routinely form in very high concentration halos. 

The location of UDGs relative to the SMHR was investigated by 
\cite{2022MNRAS.510.3356T} 
who developed a simple, semi-empirical model focusing on stellar feedback from globular clusters. In their model,  galaxies with large GC systems generate more feedback which expands the galaxy, forms a dark matter core  but also results in dark matter lost from the halo. Such galaxies lie above the SMHR, having a {\it lower} halo mass at a given stellar mass. The effect is stronger in earlier collapsing halos and those with higher star formation rates. 
In 
this model, UDGs are dwarf galaxies that have been puffed-up in size, possess DM cores and average-to-low DM contents. They conclude that their model naturally explains DM-deficient UDGs, like NGC1052-DF2/4 
\citep{2022Natur.605..435V}, 
but it does not explain those UDGs with overly massive halos (that we find). 


Globular clusters associated with normal galaxies, have been modelled within the EAGLE hydrodynamical simulation (e.g. 
\citealt{2015MNRAS.450.1937C}) in an extension called E-MOSAICS. In E-MOSAICS, early collapsing halos have 
higher densities and pressures in the ISM and this leads to more efficient GC formation 
(\citealt{2019MNRAS.486.3134K}). 
Indeed, in many star cluster formation models the fraction of stars that form in bound star clusters correlates with the star formation rate surface density. A recent study of dwarf galaxies by 
\cite{2023ApJ...949..116C} 
has cast some doubt on that relationship. We also note that observations hint at the opposite trend to that predicted with 
{\it lower} stellar surface densities (as traced by surface brightness) associated with higher fractions of mass in the GC system relative to the galaxy stellar mass 
(e.g.  
\citealt{2020MNRAS.492.4874F}; \citealt{2022A&A...665A.105L}).


As with the 
\cite{2022MNRAS.510.3356T} model, 
 the E-MOSAICS model would tend to predict GC-rich galaxies to collapse early and scatter to lower halo masses at a given stellar mass, not higher. 
Although yet to be extended to UDGs, this model would also predict GC-rich UDGs to be located near cluster cores since they should be early infallers (Pfeffer 2023, priv. comm.). This is not currently seen in infall diagnostic diagrams of UDGs
(\citealt{2022MNRAS.510..946G}; 
\citealt{2023MNRAS.525L..93F}). 
So while E-MOSAICS has been successful at explaining many GC system properties in giant galaxies 
(\citealt{2018MNRAS.475.4309P}), it has not yet been tested on the properties of UDGs and their GC systems.

Another simulation that predicts the progenitors of UDGs to form early, at high redshift, and with high GC to star efficiency is that of 
\cite{2021MNRAS.502..398C}.
Using Illustris-TNG100 they use a particle tagging technique as a proxy for GCs. In their simulation, UDGs are dwarf galaxies puffed-up by cluster tidal heating. They include the effects of tidal stripping of DM, stars and GCs. 
The simulation has had some successes, e.g. reproducing the observation that UDGs have around twice the GC mass to stellar mass ratio of normal galaxies. Interestingly, 
they predict that UDGs are offset from the standard GC system mass -- halo mass relation to higher halo masses by 0.4 dex, at a given GC system mass. For GC number, their simulation would predict that 20 GCs corresponds to a halo mass of log M$_h$ $\sim$11.4 (compared to log M$_{h}$ = 11.0 if UDGs obey the GNHR of BF20) and hence occupy very  massive halos. Their 
GC mass -- halo mass relation has significantly larger scatter than that seen observationally. 
They also predict that UDGs, at a given dynamical mass, will have a {\it lower} GC system mass than normal galaxies. For our sample of GC-rich UDGs, we find the opposite trend with UDGs having {\it more} GCs at a given
dynamical mass (see Fig. 5). 


Recently, 
\cite{2023arXiv230903260D}
have also used Illustris TNG simulations to model UDGs with GCs. They used TNG50 which has higher resolution than TNG100, but only includes clusters up to Virgo-size and not Coma-size in its 51.7 Mpc cube. The approach of Doppel et al. was to tag GCs to dark matter particles and tune their numbers to the z = 0 GC mass -- halo mass relation. Thus UDGs are effectively defined to follow the GNHR. However, they do not produce any UDGs with more than 30 GCs and consequently none of their UDGs reside in overly massive halos. Their UDGs obey the standard SMHR, or scatter to lower halo masses, depending on their infall time. Interestingly, the UDGs with the highest GC counts tend to be ancient infallers (that are quenched early-on) and to have the highest halo masses. They also note, at least one UDG that deviates from the typical halo concentration -- halo mass to a much higher concentration.

A formation pathway proposed for UDGs with massive halos is the so-called `failed galaxy' scenario 
(e.g. \citealt{2018ApJ...862...82L}; \citealt{2022ApJ...927L..28D}). 
In this picture a large number of GCs, and perhaps some field stars, form but subsequent star formation is largely quenched at early times. If these galaxies formed in relatively large 
halos with relatively inefficient star formation (due to early quenching), they will 
have high M$_{h}$/M$_{\ast}$ ratios as observed for GC-rich UDGs. Their stellar bodies may be largely made-up of the stars from disrupted GCs. 
\cite{2022MNRAS.517.2231B}
provided some support for this scenario finding that the average metallicity, at a given stellar mass, for GC-rich UDGs is low ([Fe/H] $\sim$ --1.5) and consistent with the mass--metallicity relation at z = 2.2 from the models of 
\citep{2016MNRAS.456.2140M}. In the case of NGC5846\_UDG1, it is known that the stars and GCs have the same age and metallciity (\citealt{2020A&A...640A.106M}). 

A similar process has been invoked to explain the location of 
low mass dwarf galaxies in the Local Group, relative to the SMHR.  
\cite{2019MNRAS.484.1401R}
proposed that 
early quenching of star formation effectively `freezes-in' a reduced stellar mass compared with that expected for its halo mass if star formation had proceeded. 
It is perhaps better to view such galaxies as being deficient in stars rather than overly massive in terms of their halo mass.
For the GC-rich UDGs in Fig. 8, the typical halo mass based on the GNHR is 
log M$_{h}$ = 11.32. This would correspond to a  stellar mass of log M$_{\ast}$ $\sim$ 9.3 from the SMHR. Thus a typical UDG with log M$_{\ast}$ = 8, if it originally followed the SMHR, may have only formed  
$\sim$5\% of the stars that of a normal galaxy acquires,  before it was quenched. 

Local Group dwarf galaxies have the advantage of being close enough to have resolved stellar populations and SF histories (e.g. 
\citealt{2015ApJ...804..136W})
as well as kinematic profiles (e.g. 
\citealt{2016MNRAS.459.2573R}). 
An interesting case is the Fornax dwarf galaxy. 
With R$_e$ = 0.71 kpc, $\mu_{V,0}$ = 23.3 mag per sq. arcsec, a stellar mass of  M$_{\ast}$ $\sim10^7$ M$_{\odot}$ and a dynamical mass M$_{dyn}$ = 7.39 $\times$ 10$^7$ M$_{\odot}$ \citep{2010MNRAS.406.1220W}, 
it might be considered as a NUDGE.
It hosts 6 known GCs today. The presence of these GCs, the lack of a galaxy nuclear star cluster, plus multiple analyses of its kinematics and 
inner surface density, favour a cored mass profile (see 
\citealt{2019MNRAS.487.5799R} and references therein).
Its SF continued until $\sim$1.75 Gyr ago 
\citep{2015ApJ...804..136W}
suggesting a somewhat prolonged SF history providing a mechanism to heat the DM and produce the core. 

\cite{2023MNRAS.522.5638C}
identified two model galaxies in their simulation that resembled the Fornax galaxy today in terms of its stellar mass and GC count. Initially the two model galaxies grow in stellar mass, DM  and in the number of GCs it hosts. At around 10 Gyr ago, both galaxies are quenched and lose DM due to infall onto a central galaxy with subsequent tidal stripping.  
Their GC number also peaks around this time and 
then declines by a factor of $\sim$5-8. However, this decline is largely driven by tidal disruption of the GCs themselves  giving a final  count of 6-7 GCs today. For both galaxies, the contribution of GCs via mergers was minimal. The final halo mass is $\sim$10$^{10}$ M$_{\odot}$, in good agreement with the cored 
halo mass inferred by 
\cite{2019MNRAS.487.5799R}
of 2.2 $\times$ 10$^{10}$ M$_{\odot}$ and that based on 6 GCs (and applying the GNHR) of 3 $\times$ 10$^{10}$ M$_{\odot}$. Thus the GC-based halo mass agrees well with the other halo masses for this cored dwarf galaxy. 

The Fornax dwarf galaxy with its 6 GCs today has a GC system mass to stellar mass ratio of $\sim$3\%. For both of the 
\cite{2023MNRAS.522.5638C}
model galaxies this ratio was 
considerably higher at the time of first infall/quenching. Indeed, \cite{2017MNRAS.469L..63R}
suggests that at z $\sim$5 
this ratio could be as high as 50\%.
If we scale the GC system to match that of NGC5846\_UDG1 then its 54 GCs seen today could have been several hundred GCs when its SF was largely  quenched 
\citep{2023MNRAS.526.4735F}. 
The simulation of 
\cite{2023MNRAS.522.5638C}
reveals that infall and subsequent quenching of the stars is somewhat decoupled from that of the GCs, which are reduced in number due to the stochastic effects of tidal disruption. For example, in the model of 
\cite{2023arXiv230702530D}
the disruption of GCs is a strong function of metallicity.

For UDGs, the physical process for quenching is unclear. It could be gas removal by early infall into a group/cluster 
(\citealt{2016MNRAS.455.2323M}; 
\citealt{2023arXiv230903260D}),  
interaction with a filament or cosmic sheet 
(\citealt{2023MNRAS.520.2692P}), environmental quenching associated with a dense environment (\citealt{2006MNRAS.368....2D}), 
or some form of self-quenching due to feedback from intense GC formation itself 
(\citealt{2022ApJ...927L..28D}). 
In order to maintain a cored halo, some level of ongoing star formation may be required. 
Further simulations of UDGs that 
trace both the evolution of the galaxy (physical properties, star formation history etc) and that of their GC systems are needed.

\section{Conclusions}

Here we have re-examined claims that some UDGs host rich systems of globular clusters (GCs). In particular we focused on UDGs in the literature than host {\it both} more than 20 GCs and have a measured velocity dispersion from either their stars or their GC system. \\

\noindent
Our main results are the following:\\
$\bullet$~We support claims that some UDGs do indeed host more than 20 GCs, with several having resolved GCs  from HST imaging and/or radial velocities from spectroscopy confirming their nature.\\
$\bullet$~We find that the number of GCs positively correlates with the measured dynamical mass of the host UDG within its half-light radius, as is also the case for normal galaxies. 
Thus UDGs with more GCs also tend to have larger dynamical masses. \\
$\bullet$ At a given stellar mass, GC-rich UDGs have a higher dynamical mass than normal galaxies. With typical dynamical masses of 10$^9$ M$_{\odot}$ and stellar masses of 10$^8$ M$_{\odot}$, GC-rich UDGs are highly DM-dominated within the half-light radius. \\
$\bullet$ We find good agreement between halo masses derived from the GC number -- halo mass relation (GNHR) and halo masses derived from dynamical masses assuming a cored mass profile. Since a cored halo might be expected for UDGs, this supports the application of the GNHR to estimate the halo masses of UDGs. \\
$\bullet$ By favouring the conclusion that UDGs obey the GNHR and occupy cored halos, they do not, then, follow the stellar mass -- halo mass relation (SMHR). 
If however, they have cuspy halos, then they occupy extremely under-massive halos and they do not obey the GNHR. \\
$\bullet$ We suggest that GC-rich UDGs 
occupy overly massive halos. 
However, a possible interpretation (similar to that suggested for some Local Group dwarfs) is that quenching at early times largely cuts off star formation leading to a `failed galaxy'. The result is a galaxy that occupies a relatively massive halo but has only succeeded in forming a few per cent of its potential stellar mass. \\ 
$\bullet$ Further simulations are needed that accurately model GC systems in UDGs.
None can yet successfully reproduce the observed GC system properties of GC-rich UDGs. 
Whereas observational studies should focus on obtaining more dynamical masses, particularly for low stellar mass UDGs from stellar velocity dispersions. Including other dwarf galaxies that nudge up against the UDG definition in terms of size and surface brightness (which we name here as NUDGES) could also provide useful insight.  
 
\section*{Acknowledgements}

We wish to thank the referee for their comments and suggestions that have helped us to improve the paper.
We also thank A. Romanowsky, S. Danieli, L. Buzzo, A. Ferre-Mateu, L. Haacke, J. Pfeffer, J. Brodie, O. Gnedin, R. Remus, S. Kim, J. Read and M. Collins for useful discussions.
DF thanks the ARC for support via DP220101863 and DP200102574.

\section*{Data Availability}

This study made use of publicly available data in the literature. Reasonable request for data can be made to the first author.



\bibliographystyle{mnras}
\bibliography{smhr}{}

\bsp	
\label{lastpage}
\end{document}